# Turing Instabilities and Patterns Near a Hopf Bifurcation

Rui Dilão

Grupo de Dinâmica Não-Linear, Instituto Superior Técnico,

Av. Rovisco Pais, 1049-001 Lisboa, Portugal.

and

Institut des Hautes Études Scientifiques,

35, route de Chartres, 91440, Bures-sur-Yvette, France.

rui@sd.ist.utl.pt

Phone: +(351) 218417617; Fax: +(351) 218419123

**Abstract**

We derive a necessary and sufficient condition for Turing instabilities to occur in two-component systems of reaction-diffusion equations with Neumann boundary conditions. We apply this condition to reaction-diffusion systems built from vector fields with one fixed point and a supercritical Hopf bifurcation. For the Brusselator and the Ginzburg-Landau reaction-diffusion equations, we obtain the bifurcation diagrams associated with the transition between time periodic solutions and asymptotically stable solutions (Turing patterns). In two-component systems of reaction-diffusion equations, we show that the existence of Turing instabilities is neither necessary nor sufficient for the existence of Turing pattern type solutions. Turing patterns can exist on both sides of the Hopf bifurcation associated to the local vector field, and, depending on the initial conditions, time periodic and stable solutions can coexist.

**Keywords:** Turing instability, Turing patterns, Hopf bifurcation, Brusselator, Ginzburg-Landau.

# 1-Introduction

A large number of real systems are modeled by non-linear reaction-diffusion equations. These systems show dynamic features resembling wave type phenomena, propagating solitary pulses and packets of pulses, stable patterns and turbulent or chaotic space and time behavior.

For example, in the study of the dispersal of a population with a logistic type local law of growth, Kolmogorov, Petrovskii and Piscunov, [1], introduced a nonlinear reaction-diffusion equation with a solitary front type solution, propagating with finite velocity. Hodgkin and Huxley proposed a non-linear reaction-diffusion equation to describe the propagation of pulses and packets of pulses in axons, [2]. In the beginning of the fifties, Alan Turing, [3], suggested that morphogens or evocators, identified as diffusive substances in tissues, lead to the emergence of stable patterns in organisms and in the growing embryo. In chemistry, an experimental system with diffusive reacting substances and developing stable spotty patterns has also been found, [4].

Since the seminal work of Alain Turing, [3], an enormous activity emerged in the analysis of specific reaction-diffusion systems. For an account of models involving reaction-diffusion systems in physics, chemistry, biology and ecology, see for example, Cross and Hohenberg, [5], Murray, [6], Grindrod, [7], and Carbone et al. [8].

Turing constructed a one-dimensional spatial system consisting of a finite number of connected cells arranged in a ring. In each cell, chemical substances evolve in time according to a dynamic law described by a differential equation. The flow between adjacent cells is proportional to the difference of local concentrations, and the spatial system is described by a reaction-diffusion system of equations. The linear analysis of this system lead Turing to find that, in the case of two or more diffusive and reacting substances, a stable state of the local system could be destabilized by the diffusion term, inducing a symmetry breaking in the global behavior of the extended system. This effect is called Turing or diffusion-driven instability. Depending on the type and magnitude of the unstable eigenmodes of the linearized reaction-diffusion system, different asymptotic states of the non-linear system could eventually be reached. Turing implicitly conjectured that if the eigenvalue with dominant real part of all the unstable eigenmodes of the linearized system is real, an asymptotic time independent solution of the non-linear equation could eventually be reached. On the other hand, if the eigenvalue with dominant real part of all the unstable eigenmodes is complex, the solution could evolve into a time periodic spatial function or wave. In the first case, we are in the presence of a Turing instability and, in the second case, we have an oscillatory instability.

Here, we consider the nonlinear reaction-diffusion equation,

$$\frac{\partial \varphi}{\partial t} = X(\varphi) + D.\Delta \varphi \qquad (1.1)$$

where $\varphi = (\varphi_1,...,\varphi_m)$ is a $m$-dimensional vector, $D$ is the diffusion matrix, $\Delta = \sum_{i=1}^{k} \frac{\partial^2}{\partial x_i^2}$ is the Laplace operator, and $X(\varphi)$ is a $m$-dimensional vector field. Each component of the vector of solutions of (1.1) is a scalar function defined on the set $\mathbb{R}_{0+} \times [0,S]^k$, where $[0,S]^k$ is the $k$-dimensional cube with side length $S$, and $\mathbb{R}_{0+}$ stands for the non-negative real numbers. We consider zero flux or Neumann boundary conditions,

$$\left.\frac{\partial \varphi_i}{\partial x_j}\right|_{x_p=0,S} = 0 \qquad (1.2)$$

with $i = 1,...,m$, $j, p = 1,...,k$, and bounded initial data vector functions defined on $[0,S]^k$.

To the parabolic equation (1.1), we associate the ordinary differential equation, or local system,

$$\frac{d\varphi}{dt} = X(\varphi) \qquad (1.3)$$

where $X(\varphi)$ is a vector field in the $m$-dimensional phase space $\mathbb{R}^m$. We also consider the elliptic problem associated to (1.1),

$$D.\Delta \varphi + X(\varphi) = 0 \qquad (1.4)$$

with the same boundary condition (1.2).

Fixed points of the differential equation (1.3) extend naturally to $[0,S]^k$ as constant solutions of the parabolic and elliptic equations (1.1) and (1.4), with boundary condition (1.2). Non constant solutions of the elliptic equation (1.4) are time independent solutions of the parabolic equation (1.1). These particular types of solutions are Turing solutions or Turing patterns, for short. Therefore, both equations (1.3) and (1.4) contain information about the set of time independent solutions of (1.1).

Our aim here is to derive a necessary and sufficient condition for Turing instabilities to occur in two-component reaction-diffusion equations. This will be done in section 2, where the main results of the paper, Theorems 2.2 and 2.3, will be proved. In section 3, we apply these results to the Brusselator and the Ginzburg-Landau systems of reaction-diffusion equations, and we obtain the bifurcation diagrams associated with the transition between oscillatory solutions and asymptotically stable patterns, for a small perturbation of a stable or unstable steady state of the extended system. For these systems, we have systematically searched for Turing pattern type solutions, concluding that they are not associated with Turing instabilities. In the last section we summarize the main conclusions of the paper.

**2-Turing instabilities and patterns**

We consider the two-component system of reaction-diffusion equations,

$$\frac{\partial \varphi_1}{\partial t} = f(\varphi_1, \varphi_2) + D_1 \Delta \varphi_1$$
$$\frac{\partial \varphi_2}{\partial t} = g(\varphi_1, \varphi_2) + D_2 \Delta \varphi_2$$
(2.1)

where $D_1$ and $D_2$ are the diffusion coefficients of $\varphi_1$ and $\varphi_2$, respectively, and the spatial domain is the $k$-dimensional cube of side length $S$. Assuming that the functions $f$ and $g$ are polynomials in $\varphi_1$ and $\varphi_2$, the local system associated to (2.1) can have several fixed points in phase space, limit cycles, and homoclinic and heteroclinic connections between fixed points. In these cases, the asymptotic solutions of the parabolic equation (2.1) depend on the initial data and on the topology of the $\omega$-limit sets of the local system in phase space, [9] and [10].

Without loss of generality, we assume that the local system associated to (2.1) has a fixed point at $\varphi^* = (0,0)$. Linearizing (2.1) around $\varphi^* = (0,0)$, we obtain,

$$\frac{d}{dt}\begin{pmatrix} \varphi_1 \\ \varphi_2 \end{pmatrix} = \begin{pmatrix} a_{11} & a_{12} \\ a_{21} & a_{22} \end{pmatrix}\begin{pmatrix} \varphi_1 \\ \varphi_2 \end{pmatrix} + \begin{pmatrix} D_1 \Delta \varphi_1 \\ D_2 \Delta \varphi_2 \end{pmatrix}$$
(2.2)

For initial data in $L^2([0,S]^k) \cap C^2([0,S]^k)$, and obeying Neumann boundary conditions, the solutions of (2.2) have the general form,

$$\varphi_1(x_1,...,x_k,t) = \sum_{n_1,...,n_k \geq 0} c_{n_1,...,n_k}(t) \cos(\frac{2\pi n_1}{S} x_1)...\cos(\frac{2\pi n_k}{S} x_k)$$
$$\varphi_2(x_1,...,x_k,t) = \sum_{n_1,...,n_k \geq 0} d_{n_1,...,n_k}(t) \cos(\frac{2\pi n_1}{S} x_1)...\cos(\frac{2\pi n_k}{S} x_k)$$
(2.3)

where $c_{n_1,...,n_k}(t)$ and $d_{n_1,...,n_k}(t)$ are Fourier coefficients, and, for each $t$, the components of the vector valued solution $\varphi(x,t) = (\varphi_1(x_1,...,x_k,t), \varphi_2(x_1,...,x_k,t))$ belong to the Hilbert space $L^2([0,S]^k)$. The terms under the sum in (2.3) are the eigenmode solutions of (2.2), and are indexed by a $k$-tuple of non-negative integers. Introducing (2.3) into (2.2), we obtain the infinite system of ordinary differential equations,

$$\frac{d}{dt}\begin{pmatrix} c_{n_1,\ldots,n_k} \\ d_{n_1,\ldots,n_k} \end{pmatrix} = \begin{pmatrix} a_{11} - 4D_1\frac{\pi^2}{S^2}(n_1^2+\ldots+n_k^2) & a_{12} \\ a_{21} & a_{22} - 4D_2\frac{\pi^2}{S^2}(n_1^2+\ldots+n_k^2) \end{pmatrix}\begin{pmatrix} c_{n_1,\ldots,n_k} \\ d_{n_1,\ldots,n_k} \end{pmatrix}$$
$$:= J_{n_1,\ldots,n_k}\begin{pmatrix} c_{n_1,\ldots,n_k} \\ d_{n_1,\ldots,n_k} \end{pmatrix} \quad (2.4)$$

with $(n_1,\ldots,n_k) \geq (0,\ldots,0)$. In the following, we say that, $(n_1,\ldots,n_k) \geq (m_1,\ldots,m_k)$, if $\sum_{i=1}^{k} n_i^2 \geq \sum_{i=1}^{k} m_i^2$.

For each non-negative $k$-tuple of integers $(n_1,\ldots,n_k)$, the stability of the eigenmode solutions of (2.2) around the zero fixed point is determined by the eigenvalues of the matrix $J_{n_1,\ldots,n_k}$. The matrix $J_{n_1,\ldots,n_k}$ has trace and determinant given by,

$$TrJ_{n_1,\ldots,n_k} = TrJ_{0,\ldots,0} - 4(D_1+D_2)\frac{\pi^2}{S^2}(n_1^2+\ldots+n_k^2)$$
$$DetJ_{n_1,\ldots,n_k} = DetJ_{0,\ldots,0} - 4(a_{11}D_2+a_{22}D_1)\frac{\pi^2}{S^2}(n_1^2+\ldots+n_k^2) + 16D_1D_2\frac{\pi^4}{S^4}(n_1^2+\ldots+n_k^2)^2 \quad (2.5)$$

Writing the eigenvalues of $J_{n_1,\ldots,n_k}$ as a function of the trace and determinant, we obtain,

$$\lambda_{n_1,\ldots,n_k}^{+-} = \frac{1}{2}\left(TrJ_{n_1,\ldots,n_k} \pm \sqrt{(TrJ_{n_1,\ldots,n_k})^2 - 4DetJ_{n_1,\ldots,n_k}}\right) \quad (2.6)$$

By (2.5), for every $(n_1,\ldots,n_k) \geq (0,\ldots,0)$, the real and imaginary parts of (2.6) are bounded from above, and we can define the number, [11],

$$\Lambda = \max\{\text{Re}(\lambda_{n_1,\ldots,n_k}^{+-}) : (n_1,\ldots,n_k) \geq (0,\ldots,0)\} \quad (2.7)$$

The number $\Lambda$ is the upper bound of the spectral abscissas of the set of matrices $\{J_{n_1,\ldots,n_k} : (n_1,\ldots,n_k) \geq (0,\ldots,0)\}$.

If the zero fixed point of the linear system (2.4) is asymptotically stable, then, for every $(n_1,\ldots,n_k) \geq (0,\ldots,0)$, we have $TrJ_{n_1,\ldots,n_k} < 0$ and $DetJ_{n_1,\ldots,n_k} > 0$. Under this condition, for any initial condition in $L^2([0,S]^k) \cap C^2([0,S]^k)$ and obeying Neumann boundary condition, the linear parabolic equation (2.2) has a unique asymptotic solution $\varphi = 0$.

In the nonlinear case, we take a compact domain $K$ in the phase space of the local system associated to (2.1), and containing the fixed point $\varphi^* = (0,0)$. We denote by $F_t$ the flow defined by the local system associated to (2.1). If, $F_t K \subseteq K$, for every $t \geq 0$, and for differentiable initial data with values in $K$, then system (2.1) has a solution defined for every $t \geq 0$, [9, pp. 297, Proposition 4.4]. Therefore, to any local system with a positively invariant set in phase space, is associated a well posed reaction-diffusion Cauchy problem.

**Definition 2.1** (Turing instability): *Near the fixed point $\varphi^* = (0,0)$, the reaction-diffusion system (2.1) has a Turing or a diffusion-driven instability of order $(n_1,...,n_k) \geq (0,...,0)$, if the upper bound of the spectral abscissas of the set of matrices $\{J_{n_1,...,n_k} : (n_1,...,n_k) \geq (0,...,0)\}$ is positive and coincides with one of the eigenvalues of the matrix $J_{n_1,...,n_k}$, $\lambda^+_{n_1,...,n_k} = \Lambda > 0$.*

Analogously, we say that the reaction-diffusion equation (2.1) has an oscillatory instability, if there exists some $(n_1,...,n_k) \geq (0,...,0)$ such that, $\text{Re}(\lambda^+_{n_1,...,n_k}) = \Lambda > 0$, with, $\text{Im}(\lambda^+_{n_1,...,n_k}) \neq 0$.

Depending on the sign of $\Lambda$, a constant solution of the reaction-diffusion equation (2.1) can be Turing unstable, oscillatory unstable or stable. If, $k = 1$, near the fixed point $(0,0)$, the reaction-diffusion equation (2.1) has at most one Turing instability of some order $n \geq 0$. If, $k > 1$, we can have several Turing instabilities with different orders. For example, if the reaction-diffusion equation (2.1) has a Turing instability of some order $(n_1,....,n_k)$, and $(m_1,....,m_k)$ is such that, $\sum_{i=1}^{k} n_i^2 = \sum_{i=1}^{k} m_i^2$, with $(m_1,....,m_k) \neq (n_1,....,n_k)$, then, by (2.5) and (2.6), $\lambda^+_{n_1,...,n_k} = \lambda^+_{m_1,...,m_k} = \Lambda > 0$. Around different fixed points of the local system, a reaction-diffusion system can have several Turing instabilities of different orders.

As we shall see in Theorem 2.3, when one of the diffusion coefficients is zero, it is possible to have Turing instabilities of infinite order.

The necessary and sufficient condition for the existence of Turing instabilities around the zero fixed point of a reaction-diffusion system, as a function of the diffusion coefficients and parameters of the local system, can now be analyzed.

**Theorem 2.2:** *We consider the reaction-diffusion system (2.1), and the associated linear system (2.2), with $DetJ_{0,...,0} \neq 0$, $D_1 > 0$, $D_2 > 0$, and Neumann boundary conditions in a $k$-dimensional cube of side length $S$. If the reaction-diffusion equation (2.1) has a Turing instability of some order $(n_1,...,n_k) \geq (0,...,0)$, then, one of the following conditions hold:*

*a)* $\alpha \leq 0$, $\delta > 0$, $TrJ_{0,...,0} \leq 0$ and $DetJ_{0,...,0} < 0$.

*b)* $\alpha \leq 0$, $\delta > 0$, $TrJ_{0,...,0} > 0$ and $4DetJ_{0,...,0} \leq (TrJ_{0,...,0})^2$.

*c)* $\alpha > 0$, $\delta > 0$, $TrJ_{0,...,0} \leq 0$ and $DetJ_{0,...,0} < 0$.

*d)* $\alpha > 0$, $\delta > 0$, $TrJ_{0,...,0} > 0$ and $4DetJ_{0,...,0} \leq (TrJ_{0,...,0})^2$.

*e)* $\alpha > 0$, $\delta > 0$, $TrJ_{0,...,0} \leq 0$ and $0 < DetJ_{0,...,0} < \alpha^2/4\delta$.

*f)* $\alpha > 0$, $\delta > 0$, $0 < TrJ_{0,...,0} \leq \alpha/2\delta$ and $(TrJ_{0,...,0})^2 < 4DetJ_{0,...,0} < \alpha^2/\delta - 2TrJ_{0,...,0}$.

*where* $\alpha = (a_{11}D_2 + a_{22}D_1)/(D_1+D_2)$, $\delta = D_1D_2/(D_1+D_2)^2$ *and* $\varepsilon = 4\pi^2(D_1+D_2)/S^2$. *Conditions a) to d) are also sufficient, and, for a) and b), the zero eigenmode is Turing unstable. Under the necessary condition e), there exists a Turing instability of some order, $(n_1,...,n_k) > (0,...,0)$, only if there exist integers $p_1,...,p_k$, with $\sum_{i=1}^{k} p_i^2 > 0$, such that,*

$$\alpha/2\varepsilon\delta - \sqrt{\alpha^2 - 4\delta DetJ_{0,...,0}}/2\varepsilon\delta < \sum_{i=1}^{k} p_i^2 < \alpha/2\varepsilon\delta + \sqrt{\alpha^2 - 4\delta DetJ_{0,...,0}}/2\varepsilon\delta$$

*Under the necessary condition f), there exists a Turing instability of some order, $(n_1,...,n_k) > (0,...,0)$, only if there exist integers $p_1,...,p_k$, with $\sum_{i=1}^{k} p_i^2 > 0$, such that,*

$$\sum_{i=1}^{k} p_i^2 > \alpha/2\varepsilon\delta - \sqrt{\alpha^2 - 4\delta DetJ_{0,...,0} - 2\delta TrJ_{0,...,0}}/2\varepsilon\delta$$

$$\sum_{i=1}^{k} p_i^2 < \alpha/2\varepsilon\delta + \sqrt{\alpha^2 - 4\delta DetJ_{0,...,0} - 2\delta TrJ_{0,...,0}}/2\varepsilon\delta$$

*Proof:* The proof of the theorem follows in four steps. In the first step, we embed the traces and determinants of the family of matrices $J_{n_1,...,n_k}$ in a continuous curve in a trace-determinant two-dimensional Euclidean space. Based on this construction, in the second step, we derive the necessary and sufficient condition for the existence of a Turing instability for $\alpha = (a_{11}D_2 + a_{22}D_1)/(D_1+D_2) \leq 0$ (cases a) and b)). In the third step, we derive a necessary and sufficient condition for $\alpha = (a_{11}D_2 + a_{22}D_1)/(D_1 + D_2) > 0$, obtaining cases c) and d). We also, derive simple necessary conditions that will lead to cases e) and f). In the fourth step, we improve the necessary conditions for cases e) and f) and then, the sufficient condition is derived.

*Step1:* Eliminating the integer variable $(n_1^2 + ... + n_k^2)$ from (2.5), we obtain,

$$DetJ_{n_1,...,n_k} = DetJ_{0,...,0} + \frac{(a_{11}D_2 + a_{22}D_1)}{(D_1 + D_2)}(TrJ_{n_1,...,n_k} - TrJ_{0,...,0}) + \frac{D_1 D_2}{(D_1 + D_2)^2}(TrJ_{n_1,...,n_k} - TrJ_{0,...,0})^2$$

Introducing the new variables, $y = DetJ_{n_1,...,n_k}$ and $x = TrJ_{n_1,...,n_k}$, and extending $x$ and $y$ to the continuum, the above equality defines the function,

$$y(x) = DetJ_{0,...,0} + \alpha(x - TrJ_{0,...,0}) + \delta(x - TrJ_{0,...,0})^2 \qquad (2.8)$$

where $\alpha = (a_{11}D_2 + a_{22}D_1)/(D_1 + D_2)$, $\delta = D_1 D_2 /(D_1 + D_2)^2$ and $x \leq TrJ_{0,...,0}$. The graph of the function $y(x)$, for $x \leq TrJ_{0,...,0}$, is the set $G = \{(x, y(x)) \in \mathbb{R}^2 : x \leq TrJ_{0,...,0}\}$.

Assuming that, $D_1 > 0$ and $D_2 > 0$, by (2.8), if, $\alpha \leq 0$, $y(x)$ is a monotonic decreasing function in the left neighborhood of $x = TrJ_{0,...,0}$. If, $\alpha > 0$, $y(x)$ is monotonic and increasing in the left neighborhood of $x = TrJ_{0,...,0}$, and has a minimum at some point $x < TrJ_{0,...,0}$. The curvature of the function $y(x)$ is $2\delta$, and, by the definition of $\delta$, we have $\delta \leq 1/4$.

By (2.6), an eigenmode solution of (2.2), indexed by the integers $(n_1,...,n_k) \geq (0,...,0)$, is unstable with real eigenvalues if, and only if, $TrJ_{n_1,...,n_k} \leq 0$ and $DetJ_{n_1,...,n_k} < 0$, or, $TrJ_{n_1,...,n_k} > 0$ and $DetJ_{n_1,...,n_k} \leq (TrJ_{n_1,...,n_k})^2/4$. Therefore, if the zero fixed point of the reaction-diffusion equation (2.1) has a Turing instability of some order $(n_1,...,n_k) \geq (0,....,0)$, the point $(TrJ_{n_1,...,n_k}, DetJ_{n_1,...,n_k}) \in \mathbb{R}^2$ belongs to one of the sets $T_1 = \{(x,y) \in \mathbb{R}^2 : x \leq 0, y < 0\}$ or $T_2 = \{(x,y) \in \mathbb{R}^2 : x > 0, y \leq x^2/4\}$, and necessarily, one of the intersections $T_1 \cap G$ and $T_2 \cap G$ is not empty, Figure 1.

Analogously, the real parts of the eigenvalues $\lambda^+_{n_1,...,n_k}$ are embedded into the function,

$$\lambda(x,y) = \begin{cases} \dfrac{1}{2}x, & y > x^2/4 \\ \dfrac{1}{2}\left(x + \sqrt{x^2 - 4y}\right), & y \le x^2/4 \end{cases} \qquad (2.9)$$

and if, $y(TrJ_{n_1,\ldots,n_k}) \le \left(TrJ_{n_1,\ldots,n_k}\right)^2/4$, then, $\lambda^+_{n_1,\ldots,n_k} = \lambda(TrJ_{n_1,\ldots,n_k}, y(TrJ_{n_1,\ldots,n_k}))$. The function $\lambda(x,y)$ takes negative values for $(x,y)$ in the second quadrant of the trace-determinant space. The level sets of $\lambda(x,y)$ are piecewise linear continuous curves, and an easy calculation shows that each level curve is given by,

$$\gamma_c = \{(x,y) \in \mathbb{R}^2 : x = 2c, y \ge c^2\} \cup \{(x,y) \in \mathbb{R}^2 : x < 2c, y = h_c(x) = cx - c^2\} \qquad (2.10)$$

where, $\lambda(x,y) = c \ge 0$. The curve $\gamma_c$ is piecewise linear and continuous, as shown in Figure 1.

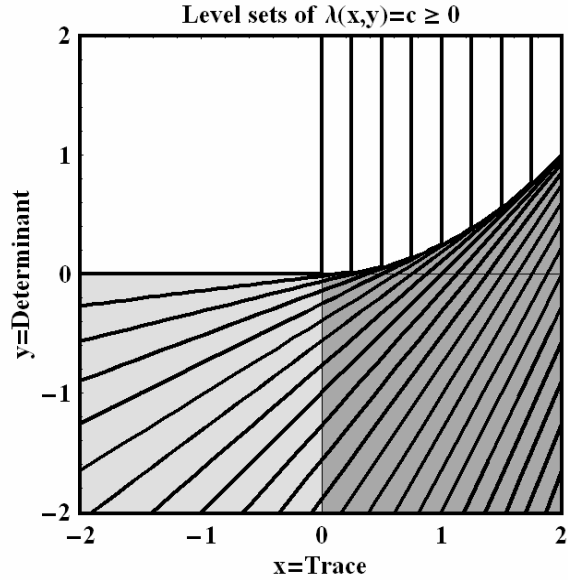

**Figure 1:** Instability sets $T_1$ and $T_2$ in the trace-determinant Euclidean space, and level sets $\gamma_c$ of the function $\lambda(x,y)$, defined in (2.9). The sets $T_1$ and $T_2$ are shown in light- and dark-grey, respectively.

*Step2:* We first consider that, $\alpha \le 0$, $\delta > 0$, and the zero fixed point of equation (2.1) has a Turing instability of some order $(n_1,\ldots,n_k) \ge (0,\ldots,0)$. As, $y(x)$ is increasing, for decreasing $x$ from $TrJ_{0,\ldots,0}$, if, $TrJ_{0,\ldots,0} \le 0$ and $DetJ_{0,\ldots,0} < 0$, then $G \cap T_1 \ne \emptyset$. So, the existence of a Turing instability implies that,

i) $\alpha \le 0$, $\delta > 0$, $TrJ_{0,\ldots,0} \le 0$ and $DetJ_{0,\ldots,0} < 0$.

Under the same hypothesis, and if, $TrJ_{0,...,0} > 0$ and $DetJ_{0,...,0} \leq (TrJ_{0,...,0})^2/4$, then $G \cap T_2 \neq \emptyset$, and, we obtain the necessary condition:

ii) $\alpha \leq 0$, $\delta > 0$, $TrJ_{0,...,0} > 0$ and $DetJ_{0,...,0} \leq (TrJ_{0,...,0})^2/4$.

Both conditions i) and ii) are exclusive and, on the curve $y(x)$, the maximum of $\lambda(x, y(x))$ is attained for $x = TrJ_{0,...,0}$. In fact, by (2.10) and (2.8), $\frac{dh_c}{dx} = c > 0$, $\frac{dy}{dx} = \alpha + 2\delta(x - TrJ_{0,...,0}) < 0$, for $x < TrJ_{0,...,0}$, and the curves $h_c(x)$ and $y(x)$ intersect transversally, Figure 1. By (2.9), an easy calculation shows that $\frac{d\lambda(x, y(x))}{dx} > 0$ for $x < TrJ_{0,...,0}$, and the maximum of $\lambda(x, y(x))$ is obtained for $x = TrJ_{0,...,0}$. Therefore, if one of the conditions i) or ii) occurs, they are necessary and sufficient for the existence of a Turing instability, and we obtain cases a) and b) of the theorem. In these cases, as $\lambda(x, y(x))$ is monotonic and increasing as $x$ decreases from $x = TrJ_{0,...,0}$, the zero eigenmode is Turing unstable.

*Step3:* If $\alpha > 0$, the function $y(x)$ has a minimum for $x < TrJ_{0,...,0}$. If, $TrJ_{0,...,0} \leq 0$ and $DetJ_{0,...,0} < 0$, or, $TrJ_{0,...,0} > 0$ and $4DetJ_{0,...,0} \leq (TrJ_{0,...,0})^2$, the zero eigenmode is unstable with a positive eigenvalue, and we consider the two cases:

iii) $\alpha > 0$, $\delta > 0$, $TrJ_{0,...,0} \leq 0$ and $DetJ_{0,...,0} < 0$.

iv) $\alpha > 0$, $\delta > 0$, $TrJ_{0,...,0} > 0$ and $4DetJ_{0,...,0} \leq (TrJ_{0,...,0})^2$.

In both cases, $\lambda(x, y(x))$ increases or decreases for decreasing values of $x$ from $x = TrJ_{0,...,0}$, Figure 1. If $\lambda(x, y(x))$ decreases, for decreasing $x$ from $x = TrJ_{0,...,0}$, the zero eigenmode is Turing unstable, and iii) and iv) are also sufficient. If $\lambda(x, y(x))$ increases, for decreasing $x$ from $x = TrJ_{0,...,0}$, due to the convexity of $y(x)$, there exists a constant $c$ such that the graph of $y(x)$ is tangent to one of the level sets of $\lambda(x, y(x))$, Figure 1. The point of tangency is obtained solving the equation, $y'(x) = h_c'(x)$, and we obtain, $x^* = TrJ_{0,...,0} - (\alpha - c^*)/2\delta$. The constant $c^*$ is such that the equation $y(x) = h_c(x)$ has only one real root ($c^* = (2\delta DetJ_{0,...,0} - \alpha)/(4\delta - 1)$ and $x^* = TrJ_{0,...,0} + (DetJ_{0,...,0} - 2\alpha)/(4\delta - 1)$). As $y(x)$ has a minimum for $x < TrJ_{0,...,0}$, by (2.8), this minimum occurs for $\bar{x} = TrJ_{0,...,0} - \alpha/2\delta$, and $\bar{x} < x^*$. Therefore, the maximum of $\lambda(x, y(x))$ occurs in one of the sets $T_1$ or $T_2$. For $x < x^*$, $\lambda(x, y(x))$ decreases. If condition iii) is fulfilled, there exists at most one unstable eigenmode with a positive eigenvalue (the zero eigenmode), and this is sufficient for a Turing instability to occur, obtaining the case c) of the theorem.

In case iv), we have two more possible cases. In the first case, the curve $y(x)$ intersects the line $y = 0$ for non-positive values of $x$, and there exists at most one eigenmode with a real and positive eigenvalue. In this case, iv) is also sufficient. In the second case, if $y(x)$ intersects the line $y = x^2/4$ for positive values of $x$, say $x = \tilde{x} > 0$, then, $\lambda(\tilde{x}, y(\tilde{x})) < \lambda(TrJ_{0,...,0}, y(TrJ_{0,...,0}))$ and the eigenvalue of the zero eigenmode is larger than the real part of any other complex eigenvalue. Therefore, iv) is also sufficient, and we have obtained the necessary and sufficient condition d) of the theorem.

Suppose now that,

v) $\alpha > 0$, $\delta > 0$, $TrJ_{0,...,0} \leq 0$ and $DetJ_{0,...,0} > 0$.

vi) $\alpha > 0$, $\delta > 0$, $TrJ_{0,...,0} > 0$ and $4DetJ_{0,...,0} > (TrJ_{0,...,0})^2$.

As $y(x)$ has a minimum for $x < TrJ_{0,...,0}$, by (2.8), this minimum occurs for, $\bar{x} = TrJ_{0,...,0} - \alpha/2\delta$, and, $y(\bar{x}) = DetJ_{0,...,0} - \alpha^2/4\delta$. In case v), and assuming the existence of a Turing instability, we necessarily have $y(\bar{x}) < 0$, implying that, $G \cap T_1 \neq \emptyset$. Then, case v), with the additional inequality $y(\bar{x}) < 0$, is necessary for the existence of a Turing instability:

v') $\alpha > 0$, $\delta > 0$, $TrJ_{0,...,0} \leq 0$ and $0 < DetJ_{0,...,0} < \alpha^2/4\delta$.

In case vi), the existence of a Turing instability implies the two additional exclusive conditions: If, $\bar{x} \leq 0$, then $y(\bar{x}) < 0$, or, if, $\bar{x} > 0$, then $y(\bar{x}) \leq \bar{x}^2/4$. So, the case vi) splits into:

vi') $\alpha > 0$, $\delta > 0$, $0 < TrJ_{0,...,0} \leq \alpha/2\delta$ and $(TrJ_{0,...,0})^2 < 4DetJ_{0,...,0} < \alpha^2/\delta$.

vi'') $\alpha > 0$, $\delta > 0$, $TrJ_{0,...,0} > \alpha/2\delta$ and $(TrJ_{0,...,0})^2 < 4DetJ_{0,...,0} \leq (TrJ_{0,...,0} - \alpha/2\delta)^2 + \alpha^2/\delta$.

But the condition vi'') leads to a contradiction. In fact, as $(TrJ_{0,...,0})^2 < (TrJ_{0,...,0} - \alpha/2\delta)^2 + \alpha^2/\delta$, it implies that $TrJ_{0,...,0} < \alpha + \alpha/4\delta$. But as $TrJ_{0,...,0} > \alpha/2\delta$, $\alpha/2\delta < \alpha + \alpha/4\delta$, implying that $4\delta > 1$, a contradiction is obtained (see the Step 1 of the proof).

Therefore, we have shown that, in the cases v') and vi'), the function $y(x)$ has a negative minimum for $x = \bar{x} \leq 0$, and this is necessary for the existence of a Turing instability.

*Step4:* We now improve the necessary conditions v') and vi') and find sufficient conditions. Solving the equation, $y(x) = 0$, we obtain the roots,

$$x^- = TrJ_{0,...,0} - \alpha/2\delta - \sqrt{\alpha^2 - 4\delta DetJ_{0,...,0}}/2\delta$$
$$x^+ = TrJ_{0,...,0} - \alpha/2\delta + \sqrt{\alpha^2 - 4\delta DetJ_{0,...,0}}/2\delta$$

By (2.5), we make,

$$TrJ_{0,...,0} - 4(D_1 + D_2)\frac{\pi^2}{S^2}(n_1^2 + ... + n_k^2) = x^-$$
$$TrJ_{0,...,0} - 4(D_1 + D_2)\frac{\pi^2}{S^2}(m_1^2 + ... + m_k^2) = x^+$$

and solving these equations in order to $(n_1^2 + ... + n_k^2)$ and $(m_1^2 + ... + m_k^2)$, we obtain,

$$(n_1^2 + ... + n_k^2) = \alpha/2\varepsilon\delta + \sqrt{\alpha^2 - 4\delta DetJ_{0,...,0}}/2\varepsilon\delta$$
$$(m_1^2 + ... + m_k^2) = \alpha/2\varepsilon\delta - \sqrt{\alpha^2 - 4\delta DetJ_{0,...,0}}/2\varepsilon\delta$$

where $\varepsilon = 4\pi^2(D_1 + D_2)/S^2$. Assuming now that v') holds, we have a Turing instability if there exist integers $p_1,...,p_k$, with $\sum_{i=1}^{k} p_i^2 > 0$, such that,

$$\alpha/2\varepsilon\delta - \sqrt{\alpha^2 - 4\delta DetJ_{0,...,0}}/2\varepsilon\delta < \sum_{i=1}^{k} p_i^2 < \alpha/2\varepsilon\delta + \sqrt{\alpha^2 - 4\delta DetJ_{0,...,0}}/2\varepsilon\delta \quad (2.11)$$

In fact, assuming v'), (2.11) implies the existence of at most one eigenmode with a positive eigenvalue, which is sufficient to ensure the existence of a Turing instability, and we obtain the case e). Note that, in this case, the zero eigenmode has an eigenvalue with a negative real part.

We consider now the case vi'). As $y(0) = DetJ_{0,...,0} - \alpha TrJ_{0,...,0} + \delta(TrJ_{0,...,0})^2$, and if the inequalities in vi') hold, it follows that $y(0) > 0$. Therefore, we are in the conditions of v'), and if we impose (2.11), we have at most one eigenmode with a real eigenvalue. However, we cannot ensure that the real part of the eigenvalue of the zero eigenmode is smaller than the eigenvalue of the eigenmode $(p_1,...,p_k)$. So, we first improve the necessary condition vi'). Imposing the additional condition, $y(x) < -TrJ_{0,...,0}/2$, after a straightforward manipulation, we obtain that, $4DetJ_{0,...,0} < \alpha^2/\delta - 2TrJ_{0,...,0}$ and,

$$\begin{aligned} x > x^- &= -\alpha/2\delta + TrJ_{0,...,0} - \sqrt{\alpha^2 - 4\delta DetJ_{0,...,0} - 2\delta TrJ_{0,...,0}}/2\delta \\ x < x^+ &= -\alpha/2\delta + TrJ_{0,...,0} + \sqrt{\alpha^2 - 4\delta DetJ_{0,...,0} - 2\delta TrJ_{0,...,0}}/2\delta \end{aligned} \quad (2.12)$$

With the first inequality, the necessary condition vi') is improved, and becomes,

vi''') $\alpha > 0$, $\delta > 0$, $0 < TrJ_{0,...,0} \leq \alpha/2\delta$ and $\left(TrJ_{0,...,0}\right)^2 < 4DetJ_{0,...,0} < \alpha^2/\delta - 2TrJ_{0,...,0}$.

Condition vi''') gives the necessary condition f). As we have done in case e), if vi''') holds, by (2.12), we have a Turing instability only if there exist integers $p_1, ..., p_k$, with $\sum_{i=1}^{k} p_i^2 > 0$, such that,

$$\begin{aligned} \sum_{i=1}^{k} p_i^2 &> \alpha/2\varepsilon\delta - \sqrt{\alpha^2 - 4\delta DetJ_{0,...,0} - 2\delta TrJ_{0,...,0}}/2\varepsilon\delta \\ \sum_{i=1}^{k} p_i^2 &< \alpha/2\varepsilon\delta + \sqrt{\alpha^2 - 4\delta DetJ_{0,...,0} - 2\delta TrJ_{0,...,0}}/2\varepsilon\delta \end{aligned}$$

and the theorem is proved. ∎

If $D_1 > 0$ and $D_2 > 0$, Theorem 2.2 gives a necessary and sufficient condition for the existence of Turing instability around the fixed point of a two component reaction-diffusion system.

It is also important to consider the case where one of the diffusion coefficients of the reaction-diffusion equation (2.1) is zero.

**Theorem 2.3:** *We consider the reaction-diffusion system (2.1), and the associated linear system (2.2), with $DetJ_{0,...,0} \neq 0$, and Neumann boundary conditions in a $k$-dimensional cube of side length $S$. We suppose further that one of the diffusion coefficients $D_1$ or $D_2$ is zero. Then, the reaction-diffusion equation (2.1) has a Turing instability of some order $(n_1,...,n_k) \geq (0,...,0)$, if, and only if, one of the following conditions hold:*

a) $\alpha \leq 0$, $TrJ_{0,...,0} \leq 0$ and $DetJ_{0,...,0} < 0$.

b) $\alpha \leq 0$, $TrJ_{0,...,0} > 0$ and $4DetJ_{0,...,0} \leq \left(TrJ_{0,...,0}\right)^2$.

c) $\alpha > 0$, $DetJ_{0,...,0} - \alpha TrJ_{0,...,0} \leq -\alpha^2$.

d) $\alpha > 0$, $DetJ_{0,...,0} - \alpha TrJ_{0,...,0} > -\alpha^2$, $4DetJ_{0,...,0} > (TrJ_{0,...,0})^2$ and $TrJ_{0,...,0} < 2\alpha$.

e) $\alpha > 0$, $DetJ_{0,...,0} - \alpha TrJ_{0,...,0} > -\alpha^2$ and $4DetJ_{0,...,0} \leq (TrJ_{0,...,0})^2$.

*and $\alpha$ assumes one of the following values: if, $D_1 = 0$ and $D_2 > 0$, then $\alpha = a_{11}$; if, $D_2 = 0$ and $D_1 > 0$, then $\alpha = a_{22}$. In cases a), b), c) and e), the zero eigenmode is Turing unstable. In case d), the Turing instability is of infinite order.*

*Proof:* The proof of this theorem follows the same line of reasoning as the proof of Theorem 2.2. If one of the diffusion coefficients is zero, (2.8) simplifies to,

$$y(x) = DetJ_{0,...,0} + \alpha(x - TrJ_{0,...,0})$$

and $\alpha$ assumes one of the following values: if, $D_1 = 0$ and $D_2 > 0$, $\alpha = a_{11}$; if, $D_2 = 0$ and $D_1 > 0$, $\alpha = a_{22}$. If, $\alpha \leq 0$, $y(x)$ is a linear and decreasing. If, $\alpha > 0$, $y(x)$ is a linear and increasing.

In the first case, $\alpha \leq 0$, and the function $y(x)$ intersects the levels sets of $\lambda(x, y(x))$ transversally, Figure 1, if, and only if, one of the following exclusive conditions hold:

i) $\alpha \leq 0$, $TrJ_{0,...,0} \leq 0$ and $DetJ_{0,...,0} < 0$.

ii) $\alpha \leq 0$, $TrJ_{0,...,0} > 0$ and $4DetJ_{0,...,0} \leq (TrJ_{0,...,0})^2$.

In cases i) and ii), the reaction-diffusion system has a Turing instability of order $(0,...,0)$, and the cases a) and b) of the theorem follow.

If, $\alpha > 0$, $y(x)$ is linear increasing, and, for sufficient low values of $x$, $y(x)$ intersects the set $T_1 = \{(x, y) \in \mathbb{R}^2 : x \leq 0, y < 0\}$. Therefore, $\alpha > 0$ is a necessary condition for the existence of a Turing instability. By (2.10), there exists a line $h_c(x) = cx - c^2$, with $c = \alpha$, parallel to $y(x)$, Figure 1. So, let us consider the two cases, $y(0) \leq h_\alpha(0)$ and $y(0) > h_\alpha(0)$, which imply,

iii) $DetJ_{0,...,0} - \alpha TrJ_{0,...,0} \leq -\alpha^2$.

iv) $DetJ_{0,...,0} - \alpha TrJ_{0,...,0} > -\alpha^2$.

For case iii), all the eigenvalues $\lambda^+_{n_1,...,n_k}$ are real, and we necessarily have a Turing instability. In this case, the zero eigenmode has the largest eigenvalue. Case iii) gives the condition c) and is necessary and sufficient.

For case iv), by (2.9) and the parallelism of the lines $y(x)$ and $h_\alpha(x) = \alpha x - \alpha^2$, if, $DetJ_{0,...,0} > (TrJ_{0,...,0})^2/4$, the largest values that $\lambda(x, y(x))$ can take are $TrJ_{0,...,0}/2$ and $\alpha$. Therefore, to have a Turing instability, it is sufficient that $TrJ_{0,...,0} < 2\alpha$, and we obtain case d). The maximum value of $\lambda(x, y(x))$ is obtained in the limit $\sum_{i=1}^{k} n_i^2 \to \infty$, and the order of the Turing instability is infinite. Note that, if $TrJ_{0,...,0} \geq 2\alpha$, the zero eigenmode has a complex eigenvalue with the largest positive real part.

Assuming iv), if, $DetJ_{0,...,0} \leq (TrJ_{0,...,0})^2/4$, the largest value that $\lambda(x, y(x))$ can take is $TrJ_{0,...,0}/2 + \sqrt{(TrJ_{0,...,0})^2 - 4DetJ_{0,...,0}}/2$, and we have a Turing instability of zero order. Then, iv) and $DetJ_{0,...,0} \leq (TrJ_{0,...,0})^2/4$ give case e). ∎

The applicability of Theorems 2.2 and 2.3 to specific cases depends on the parameterization of the local vector field associated to the reaction-diffusion equation. If the local system of the reaction-diffusion equation has a bifurcation, when we vary a parameter, the necessary and sufficient conditions given by Theorems 2.2 and 2.3 depend on the path of the eigenvalues of the fixed point on the trace-determinant Euclidean space. The above theorems contain all the possible codimension-1 bifurcation paths for two-dimensional local systems.

Another feature of Theorem 2.2 is related with the dependence of the necessary and sufficient condition on the volume of the spatial domain. In some cases, the necessary and sufficient condition depends on side length $S$ of the $k$-dimensional cube. In other cases, it depends only on the parameters of the local vector field and on the diffusion coefficients. If one of the diffusion coefficients is zero, the existence of Turing instabilities is independent of the side length of the $k$-dimensional cube.

Several necessary and several sufficient conditions for the existence of Turing instabilities in reaction-diffusion systems have been derived. This is the case of Szili and Tóth, [11], and Satnoianu et al., [12].

In applied contexts, it is sometimes suggested that the existence of Turing patterns depends on the action-inhibition features of the local dynamics, [5], [6], [11] and [13]. To be more precise, if $\partial g/\partial \varphi_1 |_{(0,0)} < 0$ and $\partial f/\partial \varphi_2 |_{(0,0)} > 0$ in (2.1), we say that, near the steady state $(0,0)$, $\varphi_2$ activates the formation of $\varphi_1$, and $\varphi_1$ inhibits the formation of $\varphi_2$. In this case, $\varphi_1$ is an inhibitor and $\varphi_2$ is an activator. Similarly, if $\partial g/\partial \varphi_1 |_{(0,0)} > 0$ and $\partial f/\partial \varphi_2 |_{(0,0)} < 0$, $\varphi_1$ is an activator and $\varphi_2$ is an inhibitor. Clearly, we can also have self-activators (autocatalysis) and self-inhibitors according the signs of the diagonal elements of the matrix $J_{0,\ldots,0}$. In this context, if, $\partial f/\partial \varphi_1 |_{(0,0)} > 0$, $\partial f/\partial \varphi_2 |_{(0,0)} < 0$ and $D_1 < D_2$, or, $\partial g/\partial \varphi_1 |_{(0,0)} < 0$, $\partial g/\partial \varphi_2 |_{(0,0)} > 0$ and $D_1 > D_2$, a reaction-diffusion system is said to obey to a principle of autocatalysis with lateral inhibition, [5] and [13]. In the next section and for specific cases, we shall compare these criteria with the results derived from Theorems 2.2 and 2.3.

**3-Systems with limit cycles**

To apply the results of the previous section, we have chosen two models with a Hopf bifurcation. One of the models, the Brusselator, [15], is the prototype of an activation-inhibition mechanism in chemistry, and has been used to mimicry the spatial dynamics of the Belousov-Zhabotinski reaction, [16]. The second model is constructed from the versal unfolding of the Hopf bifurcation, [14], and is also known as the Ginzburg-Landau model in real coordinates, [17]. From the point of view of the local dynamics, near the Hopf bifurcation, both models are topologically equivalent.

We now derive the conditions leading to Turing instabilities in the Brusselator and in the versal unfolding of the Hopf bifurcation model, as a function of the diffusion coefficients and of the parameters of the local system. Then, using numerical methods, for small perturbations of a constant solution of the reaction-diffusion system around the fixed point, we test the existence of Turing patterns. This leads to the bifurcation diagrams of the solutions of the reaction-diffusion equations.

**3.1-The Brusselator model**

We consider the Brusselator model, [15],

$$\frac{\partial U}{\partial t} = k_1 A - (k_2 B + k_4)U + k_3 U^2 V + D_1 \frac{\partial^2 U}{\partial x^2}$$
$$\frac{\partial V}{\partial t} = k_2 BU - k_3 U^2 V + D_2 \frac{\partial^2 V}{\partial x^2}$$
(3.1)

where $A$, $B$, $k_1$, $k_2$, $k_3$, $k_4$, $D_1$ and $D_2$ are positive constants. To simplify the notation, we consider a one-dimensional domain with length $S$.

The local system associated to the reaction-diffusion equation (3.1) has one fixed point with coordinates $U^* = k_1 A / k_4$ and $V^* = k_2 k_4 B / A k_1 k_3$. This fixed point has a supercritical Hopf bifurcation for $B = \frac{k_4}{k_2} + A^2 \frac{k_1^2 k_3}{k_2 k_4^2}$. If, $B \leq \frac{k_4}{k_2} + A^2 \frac{k_1^2 k_3}{k_2 k_4^2}$, the fixed point is stable. If, $B > \frac{k_4}{k_2} + A^2 \frac{k_1^2 k_3}{k_2 k_4^2}$, the fixed point is unstable and the local vector field associated to (3.1) has a limit cycle in phase space. To translate the fixed point to the origin of coordinates, we introduce the new variables, $u = U - k_1 A / k_4$ and $v = V - k_2 k_4 B / A k_1 k_3$, and we obtain,

$$\frac{\partial u}{\partial t} = (k_2 B - k_4) u + \frac{A^2 k_1^2 k_3}{k_4^2} v + \frac{B k_2 k_4}{A k_1} u^2 + \frac{2 A k_1 k_3}{k_4} uv + k_3 u^2 v + D_1 \frac{\partial^2 u}{\partial x^2}$$
$$\frac{\partial v}{\partial t} = -k_2 B u - \frac{A^2 k_1^2 k_3}{k_4^2} v - \frac{B k_2 k_4}{A k_1} u^2 - \frac{2 A k_1 k_3}{k_4} uv - k_3 u^2 v + D_2 \frac{\partial^2 v}{\partial x^2}$$

(3.2)

By (3.2), the Jacobian matrix calculated around the zero fixed point is,

$$J_0 = \begin{pmatrix} (k_2 B - k_4) & \frac{A^2 k_1^2 k_3}{k_4^2} \\ -k_2 B & -\frac{A^2 k_1^2 k_3}{k_4^2} \end{pmatrix}$$

(3.3)

and,

$$Tr J_0 = (k_2 B - k_4) - \frac{A^2 k_1^2 k_3}{k_4^2}$$
$$Det J_0 = \frac{A^2 k_1^2 k_3}{k_4^2}$$

(3.4)

To determine the regions of existence of Turing instabilities for the Brusselator model, we solve the inequalities in Theorem 2.2 in order to the diffusion coefficients and to the parameter $B$ of the local kinetics.

By (3.3) and (3.4), $Det J_0 > 0$, and cases a) and c) in Theorem 2.2 never occur. Case b) occurs if,

$$B \leq \frac{k_4}{k_2} + A^2 \frac{k_1^2 k_3}{k_2 k_4^2} \frac{D_1}{D_2}$$

$$B > \frac{k_4}{k_2} + A^2 \frac{k_1^2 k_3}{k_2 k_4^2} \quad (3.5)$$

$$B \geq \frac{k_4}{k_2} + A^2 \frac{k_1^2 k_3}{k_2 k_4^2} + 2A \frac{k_1 \sqrt{k_3}}{k_2 k_4}$$

where these inequalities result from $\alpha \leq 0$, $TrJ_{0,...,0} > 0$ and $DetJ_{0,...,0} \leq (TrJ_{0,...,0})^2/4$, respectively. Clearly, the second inequality is redundant, and the Turing instability occurs for the zero eigenmode.

By (3.3) and (3.4), the inequalities in Theorem 2.2 d) imply the two independent conditions,

$$B > \frac{k_4}{k_2} + A^2 \frac{k_1^2 k_3}{k_2 k_4^2} \frac{D_1}{D_2}$$

$$B \geq \frac{k_4}{k_2} + A^2 \frac{k_1^2 k_3}{k_2 k_4^2} + 2A \frac{k_1 \sqrt{k_3}}{k_2 k_4} \quad (3.6)$$

For the case e), we obtain,

$$B \leq \frac{k_4}{k_2} + A^2 \frac{k_1^2 k_3}{k_2 k_4^2}$$

$$B > \frac{k_4}{k_2} + A^2 \frac{k_1^2 k_3}{k_2 k_4^2} \frac{D_1}{D_2} + 2A \frac{k_1 \sqrt{k_3}}{k_2 k_4} \sqrt{\frac{D_1}{D_2}} \quad (3.7)$$

Finally, the inequalities in Theorem 2.2 f) give,

$$D_1 \leq D_2$$

$$B > \frac{k_4}{k_2} + A^2 \frac{k_1^2 k_3}{k_2 k_4^2}$$

$$B < \frac{k_4}{k_2} + A^2 \frac{k_1^2 k_3}{k_2 k_4^2} + 2A \frac{k_1 \sqrt{k_3}}{k_2 k_4} \quad (3.8)$$

$$B > \frac{k_4}{k_2} + A^2 \frac{k_1^2 k_3}{k_2 k_4^2} \frac{D_1}{D_2} + \frac{1}{k_2} \frac{D_1}{D_2} + \sqrt{2A^2 \frac{k_1^2 k_3}{k_2^2 k_4^2} \frac{D_1}{D_2} + 2A^2 \frac{k_1^2 k_3}{k_2^2 k_4^2} \frac{D_1^2}{D_2^2} + \frac{1}{k_2^2} \frac{D_1^2}{D_2^2}}$$

If one of the diffusion coefficients in the Brusselator model (3.1) is zero, by Theorem 2.3, the necessary and sufficient conditions for Turing instabilities to occur are:

$$D_1 > 0, \; D_2 = 0, \; B \geq \frac{k_4}{k_2} + A^2 \frac{k_1^2 k_3}{k_2 k_4^2} + 2A \frac{k_1 \sqrt{k_3}}{k_2 k_4} \quad (3.9)$$

or,

$$D_1 = 0, \ D_2 > 0, \ B > \frac{k_4}{k_2} \tag{3.10}$$

where inequality (3.9) results from case b) of Theorem 2.3, and (3.10) results from cases c), d) and e). Therefore, we have proved:

**Theorem 3.1:** *We consider the Brusselator system (3.1), with, $A$, $B$, $k_1$, $k_2$, $k_3$, $k_4$, $D_1$ and $D_2$ positive constants, and Neumann boundary conditions in a $k$-dimensional cube of side length $S$. If the reaction-diffusion equation (3.1) has a Turing instability of some order $(n_1,...,n_k) \geq (0,...,0)$, then one of the conditions (3.5)-(3.8) hold. Moreover, if one of the diffusion coefficients is zero, system (3.1) has a Turing instability if, and only if, one of the conditions (3.9) or (3.10) holds.*

In Figure 2, we show, in a $(D_1, B)$ diagram, the regions of existence of Turing instabilities and patterns for the Brusselator model (3.1), for one-dimensional domains of lengths $S = 3.098$ and $S = 19.365$, and calculated from the necessary and sufficient conditions of Theorem 2.2. The curve labeled "nc" is the boundary limit of the necessary condition given in the first part of Theorem 3.1.

We have searched numerically for the regions in the parameter space where Turing patterns exist. We have integrated numerically (3.1) with a benchmarked explicit numerical method, [18], where optimal convergence to the solution of the continuous system is achieved if, $\Delta t \max\{D_1, D_2\}/\Delta x^2 = 1/6$. Numerical integration for different values of $B$ and $D_1$, with $k = 1$, leads to the line separating the dark- and light-grey regions of Figure 2. In the dark-grey region, a small perturbation of the steady state of the extended system evolves into a time independent and spatially non-homogeneous steady state solution of the reaction-diffusion equation (Turing pattern). In the light-grey region, the Brusselator model has a Turing instability but the asymptotic solutions is time periodic. On both sides of the Hopf bifurcation, the Brusselator model has Turing pattern solutions.

In the bifurcation diagrams of Figure 2, Turing patterns appear if $D_1 < D_2$. By (3.3), $V$ is the activator and $U$ is the inhibitor, and the diffusion coefficient of the activator is larger than the diffusion coefficient of the inhibitor. If $D_1 > D_2$, by inequalities (3.5), we have Turing instabilities, but numerical integration of the Brusselator model (3.1) has not revealed Turing patterns. On the other hand, by (3.3), the elements of the rows of the matrix $J_0$ have the same signs, and the Brusselator model does not obey to a principle of autocatalysis with lateral inhibition, but shows Turing instability and patterns.

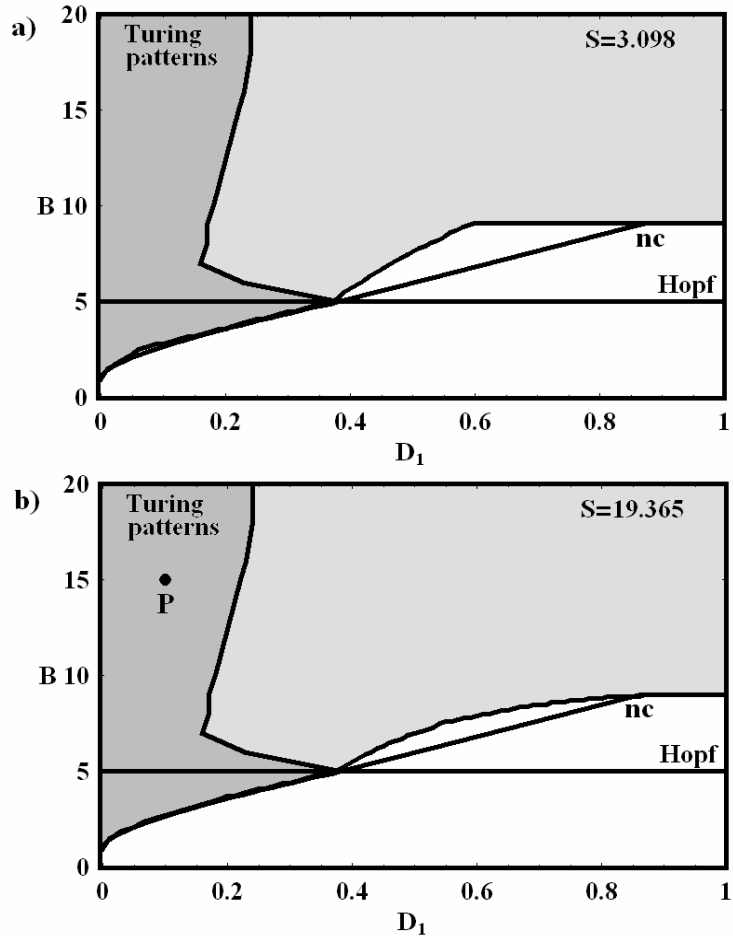

**Figure 2:** Bifurcation diagram of the solutions of the Brusselator model (3.1), for the parameter values $A=2$, $k_1=k_2=k_3=k_4=1$ and $D_2=1$, in intervals of length $S=3.098$ (a), and $S=19.365$ (b). The line $B=5$ corresponds to the supercritical Hopf bifurcation of the zero fixed point of the local system. The curve labelled "nc" is the boundary limit of the necessary condition calculated from the first part of Theorem 3.1. Both grey regions correspond to parameter values where the origin has a Turing instability, and have been calculated from Theorem 2.2. Numerical integration of equation (3.1) shows that Turing patterns only develop in the dark-grey regions. If $B>5$, in the light-grey and white regions, asymptotic solutions are constant along the spatial region and oscillatory in time with the period of the limit cycle. If $B\leq 5$, in the white regions, the asymptotic solutions are time independent and constant along the spatial region.

In Figure 3, we show the asymptotic solution of the Brusselator model for the control point P of Figure 2b), in one- and two-dimensional spatial domains. Fourier analysis of the Turing pattern shows that the number of spatial periods in the pattern is not related with the most unstable real eigenmode. In this case, the eigenmode of order $10$ has the largest eigenvalue, the eigenmodes from $0$ to $35$ are unstable with real eigenvalues, but the number of spatial periods in the Turing pattern is $7$.

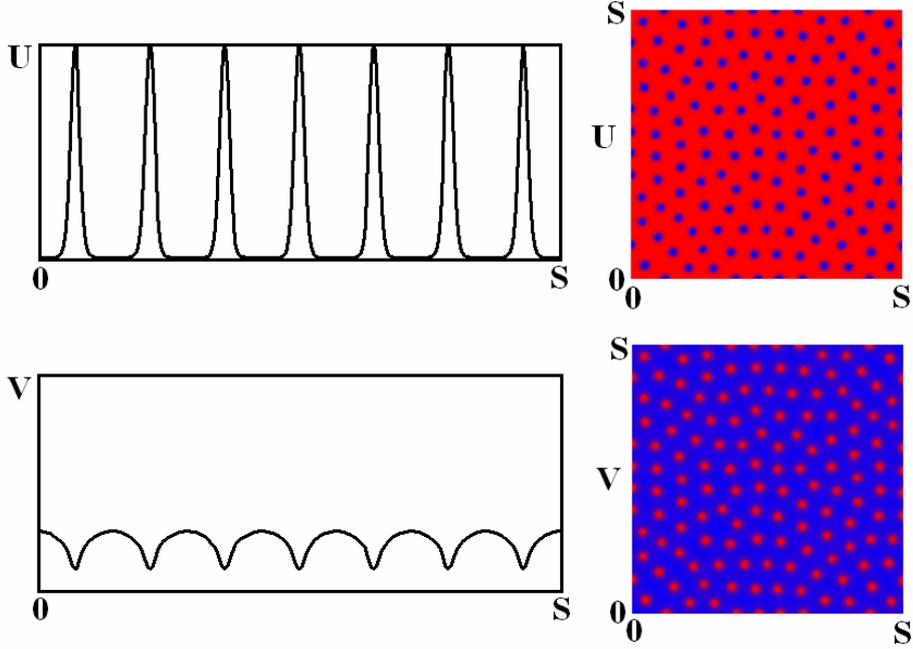

**Figure 3:** Turing pattern solution of the Brusselator reaction-diffusion equation in a one- and a two-dimensional square of side length $S$, corresponding to the control point P of Figure 2b), where $D_1 = 0.1$ and $B = 15$. The total integration time was $t = 1000$, with a time step of $\Delta t = 0.001$. We have chosen $N = 250$ lattice sites along each direction, implying that $S = N\sqrt{6\Delta t \max\{D_1, D_2\}} = N\Delta x = 19.365$. These solutions have been generated from a spatially extended random initial condition deviating slightly from the fixed point of the local system. For $B = 15$, the local system has one unstable fixed point and one limit cycle in phase space, and, for $k = 1$, the Turing instability has order $10$. For $k = 2$, we have two Turing instabilities of orders $(4,9)$ and $(9,4)$.

The simulations of Figure 3 have been generated from a spatially extended random initial condition deviating slightly from the unstable fixed point of the local system. However, for an initial condition deviating slightly from a point near the limit cycle in phase space, the asymptotic solution of the reaction-diffusion systems becomes uniform and time periodic. This shows that Turing patterns and oscillatory behavior can coexist as solutions of the same reaction-diffusion equation. A similar behavior has been reported in a reaction-diffusion system with homoclinic and a heteroclinic connection in phase space, [10].

In the Brusselator model, the existence of the Turing instability is not sufficient to ensure the existence of Turing pattern solutions of reaction-diffusion equations, and the number of spatial periods in the Turing patterns is not related with the eigenmode with the largest real eigenvalue.

The Brusselator model, defined on an interval and with Dirichlet boundary conditions, has been studied by Chow and Hale, [19]. These authors have analyzed the case where the eigenvalues of all the eigenmodes have negative real parts. Chow and Hale have shown that the boundary of stability is the union of an infinite number of curves with discontinuous derivatives. This is also apparent from Figure 2. Golubitsky and Shaffer, [20], studied the behavior of $DetJ_n$ and $TrJ_n$ as a function of $B$ and of the diffusion coefficients.

**3.2 - The normal form model**

Examples of vector fields with activation-inhibition type dynamics have, for certain parameter values, a limit cycle in phase space. Examples of systems of this type are the Brusselator, the Oregonator, [21], and the Gierer-Meinhardt [13] systems. All these system have a Hopf bifurcation and they have a stable limit cycle and an unstable fixed point, or simply a stable fixed point. Near the fixed point and for parameter values near the Hopf bifurcation, these two-component systems can be transformed, by a non-linear change of coordinates, into the local system associated with the reaction-diffusion equation,

$$\begin{cases} \dfrac{\partial \varphi_1}{\partial t} = \nu\varphi_1 - \beta\varphi_2 + (\varphi_1^2 + \varphi_2^2)(a\varphi_1 - b\varphi_2) + D_1 \dfrac{\partial^2 \varphi_1}{\partial x^2} \\ \dfrac{\partial \varphi_2}{\partial t} = \beta\varphi_1 + \nu\varphi_2 + (\varphi_1^2 + \varphi_2^2)(a\varphi_2 + b\varphi_1) + D_2 \dfrac{\partial^2 \varphi_2}{\partial x^2} \end{cases} \quad (3.11)$$

Technically, the local system associated to (3.11) is the versal unfolding of the Hopf bifurcation, [14]. In fact, when we change from Cartesian to polar coordinates, the local system associated to (3.11) transforms into,

$$\frac{dr}{dt} = r(\nu + ar^2)$$
$$\frac{d\theta}{dt} = \beta + br^2 \qquad (3.12)$$

where $\varphi_1 = r\cos(\theta)$ and $\varphi_2 = r\sin(\theta)$, and (3.12) is easily integrable by quadratures. Several attempts have been done to obtain integrable non-linear reaction-diffusion systems in order to explore pattern formation properties. For example, Kopell and Howard, [22], used a simplified form of system (3.11) to understand the mechanisms of auto-oscillations in extended systems. Other model with a parameterization similar to (3.11) is the Ginzburg-Landau reaction-diffusion system, [17].

From the phase space analysis of (3.12), it follows that, if $a < 0$, the local system has a supercritical codimension-1 Hopf bifurcation for $\nu = 0$. For $\nu \neq 0$ and $\beta \neq b\nu/a$, system (3.12) is generic, [14]. If, $\beta = b\nu/a$, (3.12) has a circumference of fixed points in phase space.

We now analyze the conditions leading to the existence of Turing instabilities in the reaction-diffusion equation (3.11). The relevant parameters to apply Theorem 2.2 are $DetJ_0 = \nu^2 + \beta^2 > 0$, $TrJ_0 = 2\nu$ and $\alpha = \nu$. If the reaction-diffusion system (3.11) has a Turing instability, by Theorem 2.2, the only conditions that give compatible inequalities are d) and f), and we have:

i) $\nu > 0$, $D_1 > 0$, $D_2 > 0$ and $\beta = 0$ (necessary and sufficient).

ii) $\nu > 0$, $D_1 > 0$, $D_2 > 0$, $\beta \neq 0$ and $\beta^2 < \dfrac{\nu^2}{4}\dfrac{(D_1 - D_2)^2}{D_1 D_2} - \nu$ (necessary).

If one of the diffusion coefficients in (3.11) is zero, by Theorem 2.3, the reaction-diffusion equation (3.11) has a Turing instability if, and only if,

iii) $\nu > 0$, $D_1 = 0$, $D_2 > 0$ and $\beta^2 \geq 0$.

iv) $\nu > 0$, $D_1 > 0$, $D_2 = 0$ and $\beta^2 \geq 0$.

By Theorem 2.2, we have tested the necessary conditions i) and ii) and we have not found Turing patterns for $D_1 = 1$ and $D_2$ small, with $\beta \neq 0$. On the other hand, the necessary and sufficient conditions iii) and iv) suggest a numerical exploration of Turing patterns when one of the diffusion coefficients is very close to zero. In Figure 4, we show a Turing pattern of the reaction-diffusion equation (3.11) for the parameter values $\nu = 1$, $\beta = -0.48$, $a = -1.0$, $b = 0.5$, $D_1 = 1$ and $D_2 = 0.001$. In this case, the reaction-diffusion equation (3.11) has no Turing instabilities, $\Lambda = 1.0$, the eigenvalue of the zero eigenmode has the largest real part, $\lambda_0^+ = 1 + i1.1$, and the reaction-diffusion equation is oscillatory unstable. This shows that the existence of a Turing instability is not a necessary condition for the existence of a Turing pattern.

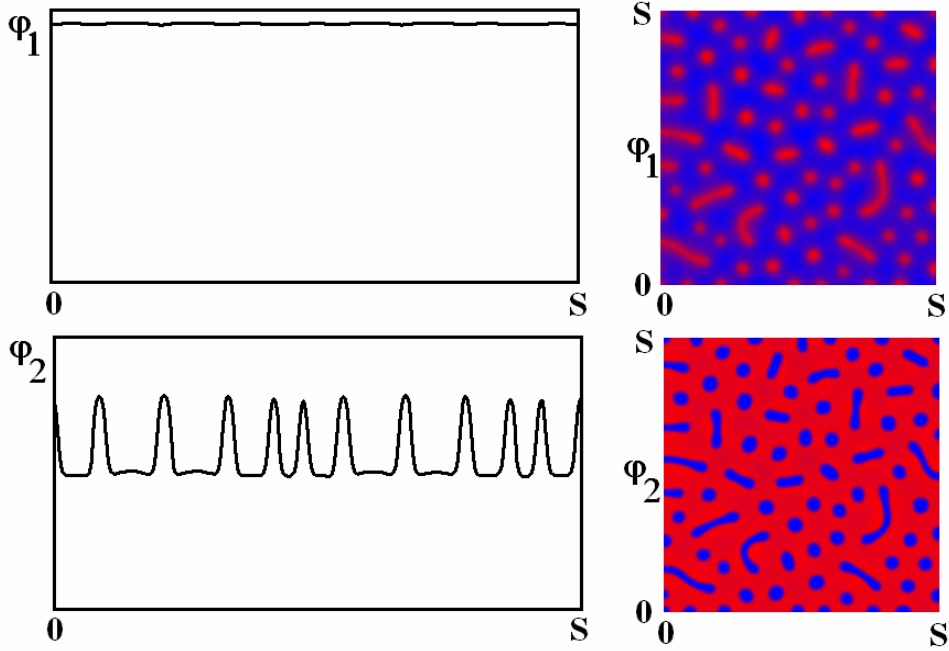

**Figure 4:** Turing patterns solutions of the normal form reaction-diffusion equation in one- and two-dimensional square region of side length $S$. The parameter values are: $v = 1$, $\beta = -0.48$, $a = -1.0$, $b = 0.5$, $D_1 = 1$ and $D_2 = 0.001$. The total integration time was $t = 1000$, with a time step of $\Delta t = 0.001$. We have chosen $N = 250$ lattice sites along each direction, implying that $S = N\sqrt{6\Delta t \max\{D_1, D_2\}} = N\Delta x = 19.365$. The solutions shown have been generated from a spatially extended random initial condition deviating slightly from the fixed point of the local system. In one and two spatial dimensions, the structure of the patterns is similar. In these Turing pattern solutions, the spatial periodicity is lost, and the patterns show quasi-periodic spatial behavior.

In this model, $\varphi_2$ is the activator and $\varphi_1$ is the inhibitor variable, and if $D_1 > D_2$, Turing patterns exist. This contrasts with the Brusselator model, where Turing patterns only appear if $D_2 > D_1$. Analyzing the Fourier coefficients of the one-dimensional Turing pattern of Figure 4, we conclude that there is a large contribution from the Fourier coefficient of the zero eigenmode, and the Fourier coefficients of eigenmodes numbers 9, 13, 14, 15 and 24 are comparable. This justifies the irregular patterning found in the stable pattern of Figure 4. The Turing pattern solutions of Figure 4 persist under small changes in the parameters $\beta$ and $D_1$. For an initial condition deviating slightly from the limit cycle, we can also have in the same system the coexistence of oscillatory and stationary solutions.

**4-Conclusions**

We have derived a necessary and sufficient condition for Turing instabilities to occur in two-component reaction-diffusion equations, with Neumann boundary conditions. We have applied the necessary and sufficient condition to the Brusselator and the Ginzburg-Landau reaction-diffusion systems, which both have a fixed point and a Hopf bifurcation when one of the parameters of the local system changes. We have obtained the bifurcation diagrams associated with the transition between oscillatory solutions and asymptotically stable patterns, for a small random perturbation of a stable or unstable steady state of the extended system.

In the case of the Brusselator model, the existence of a Turing instability is not a sufficient condition for the existence of a Turing pattern solution of the reaction-diffusion equation, and in the Ginzburg-Landau model this condition is not necessary. Therefore, in general two-component systems of reaction-diffusion equations, the Turing instability is neither necessary nor sufficient for the occurrence of Turing patterns.

In the Brusselator model, Turing patterns can exist on both sides of a Hopf bifurcation associated to the local vector field, whereas in the Ginzburg-Landau system, Turing patterns only appear when the local system has a limit cycle in phase space.

The Turing pattern type solutions of the Ginzburg-Landau system show a different patterning when compared with the Tring patterns of the Brusselator model. In the Brusselator we find a characteristic spatial period smaller than the length of the spatial domain. In the Ginzburg-Landau equation, this periodicity is lost, and we obtain a quasi-periodic pattern. Increasing the dimension of the spatial domain, this originates patterns that are a mixture of strips and spots.

As the local vector fields of the Brusselator and the Ginzburg-Landau systems are topologically equivalent near the Hopf bifurcation, in very general terms, it is difficult to understand how the emergence of collective behavior in extended systems could eventually depend on the properties of the local dynamics.

For the same reaction-diffusion system, depending on the initial conditions, time periodic and stable solutions can coexist, and we can obtain oscillatory behavior and stable patterns for the same parameter values. This type of behavior is a test for models aiming to describe experimental systems.

**Acknowledgments**: This work has been partially supported by the POCTI Project /FIS/13161/1998 (Portugal) and by *Institut des Hautes Études Scientifiques* (Bures-sur-Yvette, France).